# Coordination of Mathematics and Physical Resources by Physics Graduate Students

Ayush Gupta, Edward F. Redish, and David Hammer

*Department of Physics, University of Maryland, College Park, MD 20742.*

**Abstract.** We investigate the dynamics of how graduate students coordinate their mathematics and physics knowledge within the context of solving a homework problem for a plasma physics survey course. Students were asked to obtain the complex dielectric function for a plasma with a specified distribution function and find the roots of that expression. While all the 16 participating students obtained the dielectric function correctly in one of two equivalent expressions, roughly half of them (7 of 16) failed to compute the roots correctly. All seven took the same initial step that led them to the incorrect answer. We note a perfect correlation between the specific expression of dielectric function obtained and the student's success in solving for the roots. We analyze student responses in terms of a resources framework and suggest routes for future research.

## INTRODUCTION

Development of expertise from a novice state is a core topic of research in learning sciences [1-2]. While there are many studies that characterize novice and expert knowledge [2-5] as the two ends of the spectrum, relatively less work has been done in characterizing the intermediate stages. Expert and novice knowledge have been described within two principal frameworks: the unitary approach, in which entire concepts are considered as individual static cognitive structures [2-4], and the manifold approach, in which one's conception is determined by activation of various resources depending on the context [5-7].

In this paper we explore an intermediate level between experts and novices, from a manifold perspective. We look at how advanced graduate students coordinate very familiar mathematics knowledge in the less familiar domain of a plasma physics problem.

## TARGET PROBLEM AND SOLUTION

We analyzed students' homework for a graduate level survey course in plasma physics [8]. Of the 23 students enrolled in the course, 18 students consented to participate in the study. In this paper, we look at their solutions to a homework problem roughly four weeks into the course. The details of the question are provided in Fig. 1.

Given the velocity distribution function of the plasma constituents, the dielectric function $\varepsilon(\omega,k)$ can be derived as a function of the frequency, $\omega$, and the wave-number, $k$, by a contour integration of a function of the given velocity distribution function in the complex-$v$ plane [see reference 9 for general method]. The derivation treats the frequency $\omega$ as complex. The imaginary portion of $\omega$, however, is not just a mathematical trick to select correct boundary conditions (as in non-absorptive quantum mechanical scattering [10]). Rather, it represents absorption in the medium (damping) [9]. Alternatively, one could evaluate the integral for $\varepsilon(\omega,k)$ using the method of calculating the principal value of the integral and including the contribution due to deviation from the real axis, thus allowing for a complex $\omega$. The result can be expressed as either of the following.

$$\varepsilon(\omega,k) = 1 - \frac{\omega_p^2}{(\omega + ikv_{th})^2} \quad (1a),$$

$$\varepsilon(\omega,k) = \left(1 - \frac{\omega_p^2(\omega^2 - k^2v_{th}^2)}{(\omega^2 + k^2v_{th}^2)^2}\right) + i\frac{2\omega_p^2\omega kv_{th}}{(\omega^2 + k^2v_{th}^2)^2} \quad (1b).$$

Both expressions are correct and one can be obtained from the other by the standard procedure for rationalizing the denominator in Eq. 1(a), to bring the explicit '$i$' into the numerator. However, in this case 'rationalization of the denominator' is misleading: Although the expression in Eq. 1(b) looks like an expansion into real and imaginary parts, it is not; ω remains a complex variable. For convenience, we refer to the form of $\varepsilon(\omega,k)$ in Eq. 1(a) as the "compact-form" and that in Eq. 1(b) as "expanded-form". The compact form is directly obtained by a particular choice of the contour during the complex integration, while the principal-value method yields the expanded form. Of course, one could expand or simplify $\varepsilon(\omega,k)$ at any intermediate step. The roots of the equation $\varepsilon(\omega,k) = 0$ are obtained by setting the right side of Eq. 1(a) to zero. They are $\omega = \pm\omega_p - ikv_{th}$.

---

Obtain the dielectric function $\varepsilon(\omega,k)$ for the following three choices of one-dimensional distribution functions (a) $f_o(v) = \dfrac{n_o v_{th}}{\pi} \dfrac{1}{v_{th}^2 + v^2}$ (Omitting the other two for the purposes of this paper), where $v_{th}$ represents the typical velocity. Make graphs of the real and imaginary parts of dielectric constant for real $\omega$ and $k$ positive.
Solve for the roots of $\varepsilon=0$ in each case. Show that all give $\omega=\pm\omega_p$ in the $k=0$ limit, but have different corrections for small $k$. Explain why some roots are damped and others are not. (Hint: Distribution (a) can be solved using contour integration.)

**Figure 1.** Problem Statement (courtesy Thomas M. Antonsen, Jr.)

## ANALYSIS OF STUDENTS' SOLUTIONS

The pattern of students' success in solving for the roots of $\varepsilon(\omega,k)$ and the expression of $\varepsilon(\omega,k)$ they obtained is summarized in Table 1. Of the 18 consenting students, one did not submit the homework and another student's homework was incomplete and could not be analyzed. Of the 16 homeworks analyzed, 7 obtained $\varepsilon(\omega,k)$ in the expanded form only. These students did not write down the compact form at any intermediate step. The other 9 derived $\varepsilon(\omega,k)$ in the compact form, either directly or by simplifying the expanded form that they got at an intermediate stage. Of these 9 students, 6 drew explicit attention to both the forms (either writing both forms simultaneously or by expanding $\varepsilon(\omega,k)$ to make the plots of real and imaginary parts of $\varepsilon(\omega,k)$ for real $\omega$ and positive $k$) before solving for the roots.

We observed a 100% correlation in our data between obtaining the dielectric function in the compact form and being able to correctly solve for its roots. All students who used the compact form of dielectric function equated the entire right side of Eq. 1(a) to zero to solve for the roots correctly. It should be noted that these students did not need to treat ω as a complex variable in order to get to the correct solution: the complex roots came out just following algebraic manipulation of the equation. All students who only had access to the expanded-form expression, considered the first and second (without the '$i$' factor) terms of Eq. 1(b) as the real and imaginary parts of $\varepsilon(\omega,k)$ respectively. They solved for the roots by equating the first and second terms individually to zero (Some only set the first term to zero for computing roots).

What led students to make the erroneous assumption that the expanded form separated the real and imaginary parts of $\varepsilon(\omega,k)$? One explanation could be that the form of Eq.1(b) activated a familiar complex-algebra resource of solving an equation of the form

$$z = a + ib = 0 \qquad (2).$$

Typically, in this case $a$ and $b$ are real and the solution step involves equating $a$ and $b$ to zero individually. This in combination with the complex nature of ω not being explicit could have led students to the erroneous step. Another explanation could be that some students were simply not aware that ω could be complex and so were more likely to make the error of identifying the individual terms in Eq. 1(b) as real.

Table 1. Pattern of expression for $\varepsilon(\omega,k)$ obtained and success in solving for roots.

| Form of $\varepsilon(\omega,k)$ obtained | Solved correctly for roots | Solved incorrectly for roots |
|---|---|---|
| 1(a) only | 3 | 0 |
| 1(a) and 1(b) | 6 | 0 |
| 1(b) only | 0 | 7 |

We looked at the solutions in detail to analyze students' explicit use of ω as real or complex elsewhere in their solution. The students who drew ω/k as complex in the v-plane, or those who explicitly noted that ω can have an imaginary part, calculated it, or discussed damping based on the imaginary part of ω were listed as treating 'ω as complex,' evidence they were aware at some point of ω as complex. Those who drew $\omega/k$ on the real $v$-axis or made no explicit mention of $\omega$ as complex were listed as treating '$\omega$ as real'. The results are presented in Table 2: Whether students treated $\omega$ as complex elsewhere in their solution did not determine whether they were able to solve for the roots correctly. 71% (5 of 7) of students using the dielectric function in only the expanded form showed some awareness of $\omega$ as complex. Of this group, 3 students tried to justify their solution by stating that the imaginary part of $\omega$ is very small but 2 other students from the same group went on to explicitly compute the imaginary part of $\omega$ in order to determine damping. The 10 students who showed awareness of $\omega$ as complex were evenly divided in solving for the roots correctly and incorrectly – the division being determined by whether they had access to the compact form of the dielectric function. Of the 6 students who did not make any explicit reference to $\omega$ as complex, the majority (4 out of 6 or 66%) did solve for the roots correctly.

Table 2. Pattern of students treating $\omega$ as real or complex and success in solving for roots

|  | Treated $\omega$ as complex | Treat $\omega$ as real |
| --- | --- | --- |
| Solved correctly for roots | 5 | 4 |
| Solved incorrectly for roots | 5 | 2 |

These results seem to suggest that the specific form of the dielectric function obtained had a much greater influence on how students interpreted the terms in that expression and solved for the roots than a conception about the real/complex nature of $\omega$.

We suggest a resources-based explanation for this. The homework involves not only non-trivial calculations (algebra, complex integrations) but also coordination of newly learned physics information. The expanded form of epsilon could activate the stable resource of solving familiar equations of the form Eq. 2. It is likely that the newly learned association of the frequency, $\omega$, with a complex variable is relatively less stable and has a lower cuing priority. Also, $\omega$ as written in Eq. 1(b) does not naturally cue that it is a complex variable, just as one would be likely to interpret an equation of the form $ax+by=0$ as involving $a,b$ as real constants and $x,y$ as real variables.

Thus we posit that in the context of solving for the roots of $\varepsilon(\omega,k)$, these students fell into treating $\omega$ as real, although at other moments of the solution they understood $\omega$ to be complex, as was presupposed in the question ("why are some roots damped?") and was a part of explicit discussion in class. That is, the real/complex nature of $\omega$ is not stable in the context of this homework. Some of the students did not hesitate to consider $\omega$ as complex or even to compute the imaginary part of $\omega$, in response to the question on damping, right after solving for the roots as if $\omega$ was real.

Students who wrote the compact form of $\varepsilon(\omega,k)$ could access other methods for solving the expression $\varepsilon(\omega,k)=0$ – solving the equation $z=0$ where $z$ is not in the $a+ib$ form – leading to a separate set of algebraic steps which led them to the correct solution of the roots. This is apparent by the fact that none of those who had access to *both* forms (6 students) made the erroneous step in interpreting the terms in the dielectric function expression. These students chose to use the compact form of the dielectric function to solve for roots (perhaps simply for the relative algebraic ease of using the compact expression).

## DISCUSSION

There are a number of constraints imposed by the nature of the data. We cannot be sure that the chronology of the submitted written material represents the chronology of student thinking, or that it represents all the possibilities that the student might have considered. Analyzing submitted homework also hides from the researchers' eyes the influence of collaborative work among students or that of the professors' comments during office hours. This makes it difficult to attribute a stable conception to the students regarding whether $\omega$ is real or complex. There could also be issues of epistemological framing [6] that we have not explored. Do students see the physical significance of the frequency $\omega$ as a complex variable when they are solving the integrals or do they consider it a mathematical procedure? Future studies that observe students *in situ* solving homework problems along with follow-up interviews could shed further light on how advanced students coordinate their resources.

We would also like to clarify that we are not calling for instructionally encouraging students to use one or the other form of $\varepsilon(\omega,k)$ for similar problems. The two different expressions for $\varepsilon(\omega,k)$ only lead to slightly different paths that the students follow but did not seem to reflect any demonstrable difference in the way they conceptually treated the problem. Rather, the instructional implication is that even in advanced courses we need to pay attention to how students treat the interplay of physical concepts and mathematics.

## CONCLUSIONS

We investigate how advanced graduate students – near experts in simple complex variable manipulations – coordinate their knowledge of complex numbers for solving a problem in plasma physics – a context on which they are not so expert. Our results suggest that a manifold or resources-based framework could be used productively to understand students' solutions and their mistakes. We suggest more detailed future studies that include interviews for better characterization of their knowledge.

There are numerous studies that explore expert and novice knowledge and model them as static and unitary or as dynamic, emergent, and manifold cognitive structures. In either case, modeling the development of expertise is a central topic of research. There are obvious difficulties in monitoring the trajectories of individuals from naïveté to expertise. Indeed one cannot even predict that the starting novices will develop into experts. The alternative is to study a cross-section of the population at various stages of development. This work aims at providing some insight and motivating more detailed future research.

## ACKNOWLEDGEMENTS

We thank Thomas M. Antonsen, Jr. for letting us reproduce from the course homework assignment and study participants for providing access to their written work. We also thank the members of Physics Education Research Group at the University of Maryland for discussions. This work was supported in part by NSF grants REC 0440113, DUE 05-24987 and DUE 0524595.